\documentclass[12pt]{article}  

\def\[{\left\lbrack}
\def\]{\right\rbrack}

\def\({\left(}
\def\){\right)}
\newcommand{\be}{\begin{equation}}
\newcommand{\ee}{\end{equation}}
\newcommand{\ea}{\end{eqnarray}}
\newcommand{\ba}{\begin{eqnarray}}

\usepackage{graphicx}

\begin{document}

\title{Tunneling probability for the birth of universes with radiation, cosmological constant and an ad hoc potential}

\author{G. Oliveira-Neto and D. L. Canedo\\
Departamento de F\'{\i}sica, \\
Instituto de Ci\^{e}ncias Exatas, \\ 
Universidade Federal de Juiz de Fora,\\
CEP 36036-330 - Juiz de Fora, MG, Brazil.\\
gilneto@fisica.ufjf.br, danielcanedo.tr@hotmail.com
\and G. A. Monerat\\
Departamento de Modelagem Computacional, \\
Instituto Polit\'{e}cnico, \\
Universidade do Estado do Rio de Janeiro, \\
CEP 28.625-570, Nova Friburgo - RJ - Brazil.\\
monerat@uerj.br}

\date{\today}

\maketitle

\begin{abstract} 
In this work we study the birth of Friedmann-Lema\^{\i}tre-Robertson-Walker (FLRW) models with 
zero ($k=0$) and negative ($k=-1$) curvatures of the spatial sections. The material content of the models is composed of a
radiation perfect fluid and a positive cosmological constant. The models also have the presence of an ad
hoc potential which origin is believed to be of geometrical nature. In order to describe the birth of these
universes, we quantize them using quantum cosmology. Initially, we obtain the Wheeler-DeWitt equations and
solve them using the WKB approximation. We notice that the presence of the ad hoc potential produces a barrier
for any value of $k$. It means that we may describe the birth of the universe through a tunneling mechanism,
for any curvature of the spatial sections, not only for the usual case $k=1$. We, explicitly, compute the
tunneling probabilities for the birth of the different models of the universe and compare these tunneling 
probabilities. 
\end{abstract}



\section{Introduction}

Quantum cosmology (QC) was the first attempt to describe the Universe
as a quantum mechanical system. It uses general relativity (GR) in
order to describe the gravitational interaction between the
material constituents of the Universe. The canonical quantization
was the first method used by the physicists working in QC.
Several physicists contributed to the development of that research
area, culminating in the introduction of the Wheeler-DeWitt equation 
\cite{dewitt}, \cite{wheeler}. Another way to quantize a theory is
using the path integral method \cite{feynman,feynman1}. That
method was first discussed in connection to the quantization of GR
by C. Misner \cite{misner}. After that, many physicists contributed
to the development of that method of quantization in QC. Another
fundamental line of research in QC is the problem of interpretation.
Since one cannot use the Copenhagen interpretation of quantum mechanics
to the system composed of the entire Universe, several new interpretations
of quantum mechanics have been introduced. The first one was {\it De Broglie-Bohm}
or {\it Causal Interpretation}, first suggested by L. de Broglie 
\cite{debroglie,debroglie1,debroglie2,debroglie3,debroglie4} and later developed 
by D. Bohm \cite{bohm,bohm1}. Another important interpretation was formulated by
H. Everett, III and is known as the {\it Many Worlds Interpretation} 
\cite{everett}. A more recent interpretation of quantum mechanics that may be
used in QC is the {\it Consistent Histories} or {\it Decoherent Histories}
\cite{rbgriffiths,omnes,omnes1,omnes2,gellmann,gellmann1,gellmann2}.
For a more complete introduction of the basic concepts of QC see 
\cite{halliwell,paulo,kiefer,julio}.

One of the most interesting explanations for the regular birth of
the Universe, coming from QC, is the spontaneous {\it creation from 
nothing} \cite{grishchuk, vilenkin, vilenkin1, vilenkin2, hawking, 
linde, rubakov, vilenkin3}. In that explanation, one has to consider
the Universe as a quantum mechanical system, initially with zero size.
It is subjected to a potential barrier which confines it. In a 
FLRW quantum cosmological model, that potential
barrier is formed, most generally, due to the positive curvature of the 
spatial sections of the model and also due to the presence of a positive 
cosmological constant or a matter content that produces an accelerated 
expansion of the universe. Since, in that explanation, the Universe should 
satisfy the quantum mechanical laws, it may tunnel through the barrier and 
emerges to the right of it with a finite size. That moment is considered the 
beginning of the Universe. Therefore, the Universe starts in a regular way 
due to its finite size. Several works, in the literature, have already 
considered cosmological models where one can compute, quantitatively, the 
tunneling probability for the birth of different universes 
\cite{paulo1,acacio,germano,germano1,germano2,rocha}.

Since there are some theoretical \cite{guth,linde1} as well as observational 
\cite{george,anton} evidences that our Universe has a flat spatial geometry,
it would be interesting if we could produce a spatially flat cosmological model
which birth is described by a spontaneous {\it creation from nothing}. As
mentioned, above, the usual way to construct the potential barrier uses as one
of the fundamental ingredients the positive curvature of the spatial sections
of the universe. Therefore, one has to find a different way to produce the
barrier that the Universe has to tunnel through in order to be born. In a
recent paper some of us have introduced an ad hoc potential ($V_{ah}$), that has all
necessary properties in order to describe the regular birth of the Universe by 
the spontaneous {\it creation from nothing} \cite{germano2}. In addition to those 
properties, the universe could have positive, negative or nil curvature of the 
spatial sections. It is believed that such ad hoc potential may appear as a purely 
geometrical contribution coming from a more fundamental, geometrical, gravitational 
theory than general relativity \cite{germano2}. As mentioned in Ref. \cite{germano2}, 
$V_{ah}$ has another interesting property at the classical level. It produces a large 
class of non-singular, bounce-type solutions.

In this work we study the birth of FLRW models with zero ($k=0$) and negative ($k=-1$) curvatures of the spatial sections. 
The model with $k=1$ was studied in Ref. \cite{germano2}. Here, we are going to consider few results obtained in Ref. \cite{germano2}
in order to compare them with the new results obtained for the models with $k=0$ and $k=-1$. The material content of the models 
is composed of a radiation perfect fluid and a positive cosmological constant. The models also 
have the presence of an ad hoc potential which origin is believed to be of geometrical nature. 
In order to describe the birth of these universes, we quantize them using quantum cosmology. 
Initially, we obtain the Wheeler-DeWitt equations and solve them using the WKB approximation. 
We notice that the presence of $V_{ah}$ produces a barrier for any value of $k$. It means that 
we may describe the birth of the universe through a tunneling mechanism, for any curvature of the 
spatial sections, not only for the usual case $k=1$. We, explicitly, compute the tunneling 
probabilities for the birth of the different models of the universe and compare these tunneling 
probabilities.

In Section \ref{CM}, we obtain the Hamiltonians of the models and investigate the possible
classical solutions using phase portraits. In Section \ref{CQ}, we canonically quantize the models
and write the appropriate Wheeler-DeWitt equations. Then, we find the approximated WKB solutions to
those equation. In Section \ref{results}, we compute the quantum WKB tunneling probabilities as 
functions of the parameters: (i) the ad hoc potential parameter ($\sigma$), (ii) the cosmological 
constant ($\Lambda$), (iii) the radiation energy ($E$) and (iv) the curvature parameter ($k$). As
the final result of that section, we compare the $TP_{WKB}$'s for models with different values of $k$.
The conclusions are presented in Section \ref{CC}. In 
Appendix A, we give a detailed calculation of the fluid total hamiltonian used in this work.

\section{The Classical Model}
\label{CM}

In the present work, we want to study homogeneous and isotropic universes with constant negative and nil curvatures of the spatial sections.
Therefore, we start introducing the FLRW metric, which is the appropriate one to treat those universes,
\begin{equation} 
\label{métrica}
    ds^{2} = -N^{2}(t)dt^{2} + a^{2}(t) \Bigg{(}\frac{dr^{2}}{1 - kr^{2}} + r^{2}d\Omega^{2} \Bigg{)},
\end{equation}
where $a(t)$ is the scale factor, $k$ gives the type of constant curvature of the spatial section, $d\Omega$ is
the angular line element of a $2D$ sphere and $N(t)$ is the lapse function introduced in the ADM formalism \cite{wheeler}. The action of the
geometrical sector of the model is given by,
\begin{equation} 
\label{Ação grav}
    S = \frac{1}{2} \int_{M}d^{4}x\sqrt{-g}(R - 2\Lambda) + \int_{\partial M}d^{3}x\sqrt{h}h_{ab}K^{ab}
\end{equation}
where $R$ is the Ricci scalar, $\Lambda$ is the cosmological constant, $h_{ab}$ is the 3-metric induced on the boundary $\partial M$ of the four-dimensional space-time $M$ and $K^{ab}$ is the extrinsic curvature tensor of the boundary. We use the natural unit system where 
$\hbar = 8\pi G = c = k_{B} = 1$. After some calculations we obtain from the action Eq. (\ref{Ação grav}), with the aid of the metric coming from Eq. (\ref{métrica}), the following hamiltonian for the gravitational sector, 
\begin{equation} 
\label{Hamiltoniana_grav}
     N \mathcal{H} = -\frac{p^{2}_{a}}{12} - 3ka^{2} + \Lambda a^{4},
\end{equation}
where $p_{a}$ is the canonically conjugated momentum to $a$. Here, we are working in the conformal gauge $N = a$.
The matter content of the models is a radiation perfect fluid, which is believed to have been very important in the beginning of our universe. That perfect fluid has the following equation of state,
\begin{equation}
\label{eqstate}
p_{rad} = \frac{1}{3}\rho_{rad},
\end{equation}
where $p_{rad}$ is the radiation fluid pressure and $\rho_{rad}$ is its energy density.
In order to obtain the hamiltonian associated to that fluid, we use the Schutz variational formalism \cite{schutz,schutz1}. The starting point for that task is the following perfect fluid action \cite{julio1},
\begin{equation} 
\label{Ação fluido}
    \int_{M}d^{4}x\sqrt{-g} p_{rad}
\end{equation}
The necessary calculations in order to obtain the hamiltonian from that action Eq. (\ref{Ação fluido}), using the Schutz variational formalism, are presented in Appendix \ref{SF}.

Using Eq. (\ref{Hamiltoniana_grav}) and Eq. (\ref{hamiltoniana fluido_2}) from Appendix \ref{SF}, we may write the total hamiltonian of the model, in the conformal gauge $N = a$, as,
\begin{equation} 
\label{Hamiltoniana}
     N \mathcal{H} = -\frac{p^{2}_{a}}{12} + p_{T} - 3ka^{2} + \Lambda a^{4} + V_{ah},
\end{equation}
where $p_{a}$ and $p_{T}$ are the canonically conjugated momenta to $a$ and $T$, respectively. The variable $T$ is associated to the radiation fluid, as discussed in Appendix \ref{SF}. $V_{ah}$ is the ad hoc potential, which is defined as,
\begin{equation} 
\label{ad hoc}
    V_{ah} = - \frac{\sigma^{2} a^{4}}{(a^{3} + 1)^{2}},
\end{equation}
where $\sigma$ is a dimensionless parameter associated to the magnitude of that potential.
As discussed in Ref. \cite{germano2}, if one observes the limits of the ad hoc potential Eq.(\ref{ad hoc}), when $a$ assumes small as well as large values, one notices that it produces a barrier. In FLRW cosmological models constructed using the Ho\v{r}ava-Lifshitz gravitational theory \cite{horava,bertolami,kord,gil3,gil}, one may have, in the hamiltonian, terms similar to the asymptotic limits of $V_{ah}$, which have purely geometrical origin. Then, it is not difficult to imagine that $V_{ah}$ should come from a purely geometrical contribution of a more fundamental gravitational theory.

\begin{figure}[!htb]
\begin{minipage}[t]{0.45\textwidth}
\centering
\includegraphics[width=\linewidth]{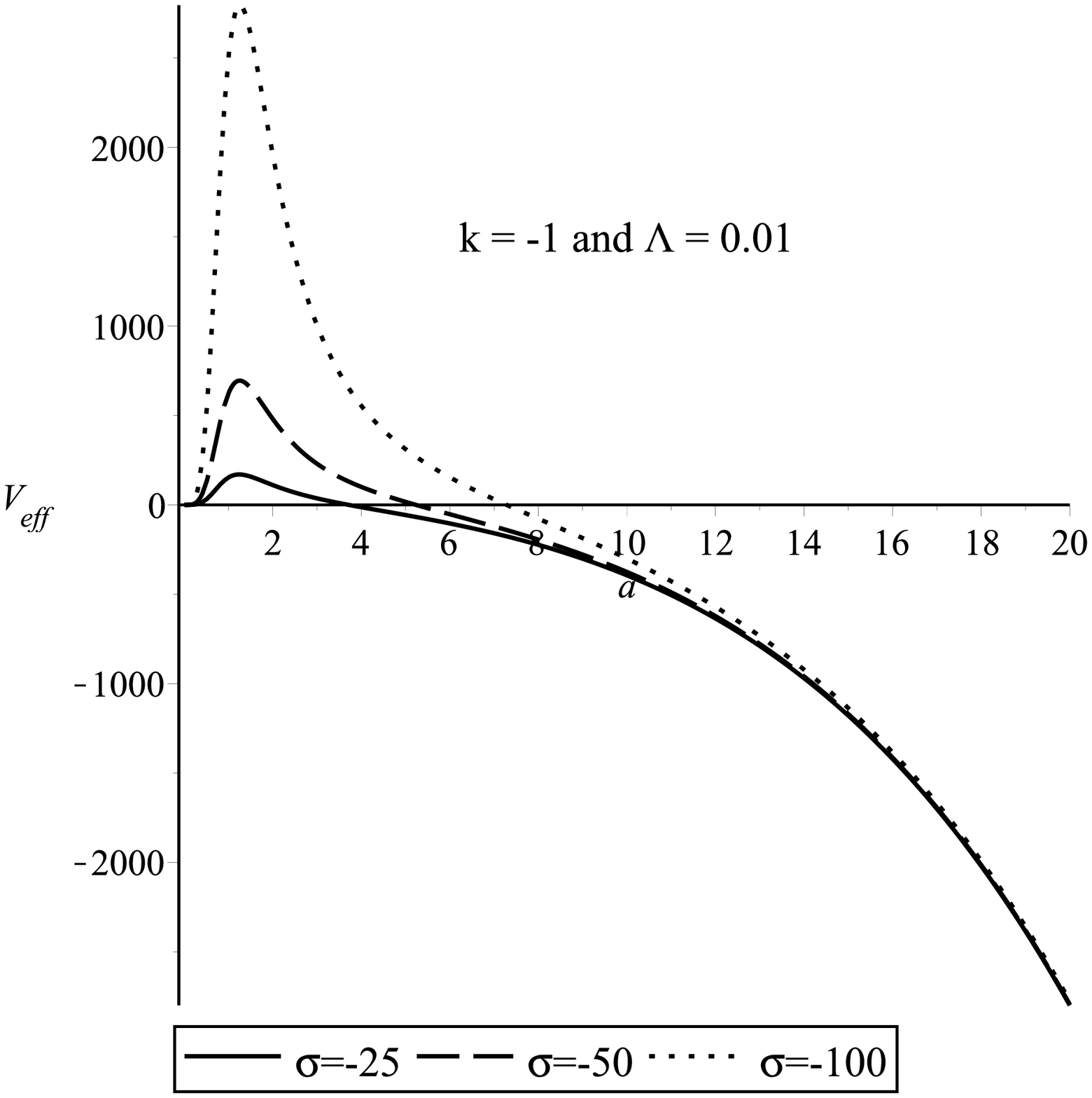}
\caption{$V_{eff}(a)$ for $k=-1$ with $\Lambda = 0.01$ and different values of $\sigma$.}
\label{Figura 1}
\end{minipage} \hfill
\begin{minipage}[t]{0.45\textwidth}
\centering
\includegraphics[width=\linewidth]{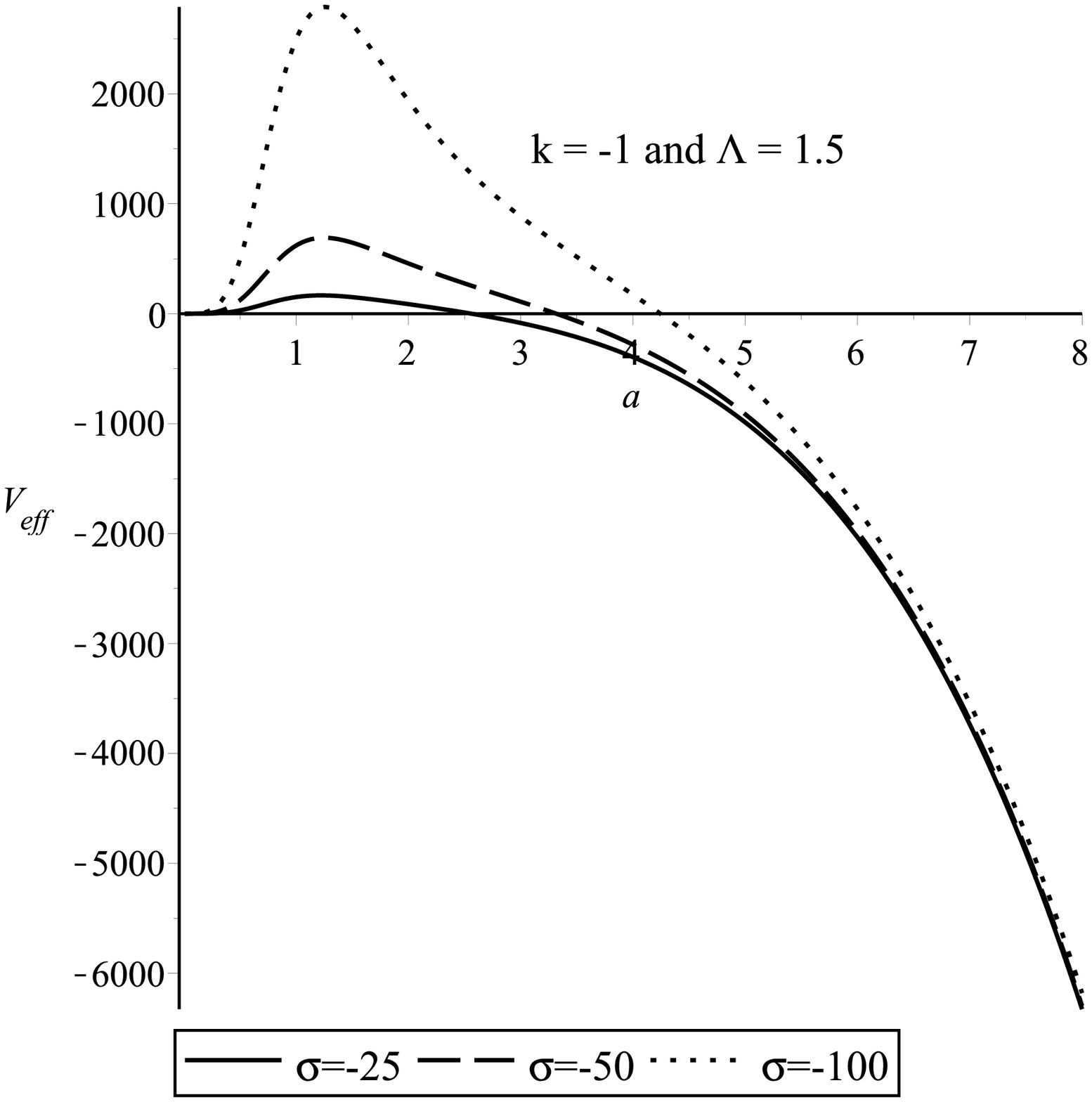}
\caption{$V_{eff}(a)$ for $k=-1$ with $\Lambda = 1.5$ and different values of $\sigma$.}
\label{Figura 2}
\end{minipage} \hfill
\end{figure}

\begin{figure}[!htb]
\begin{minipage}[t]{0.45\textwidth}
\centering
\includegraphics[width=\linewidth]{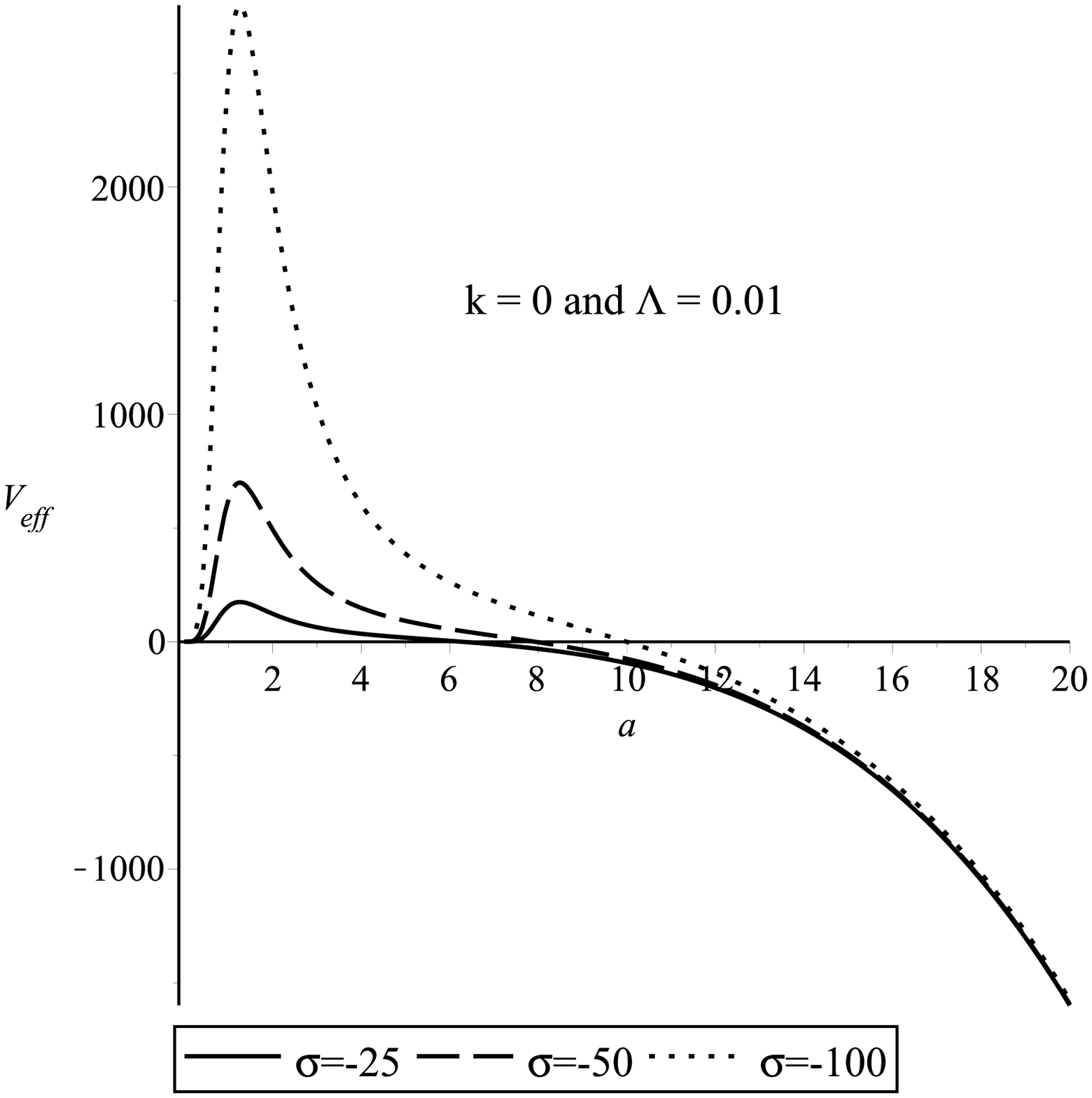}
\caption{$V_{eff}(a)$ for $k=0$ with $\Lambda = 0.01$ and different values of $\sigma$.}
\label{Figura 3}
\end{minipage} \hfill
\begin{minipage}[t]{0.45\textwidth}
\centering
\includegraphics[width=\linewidth]{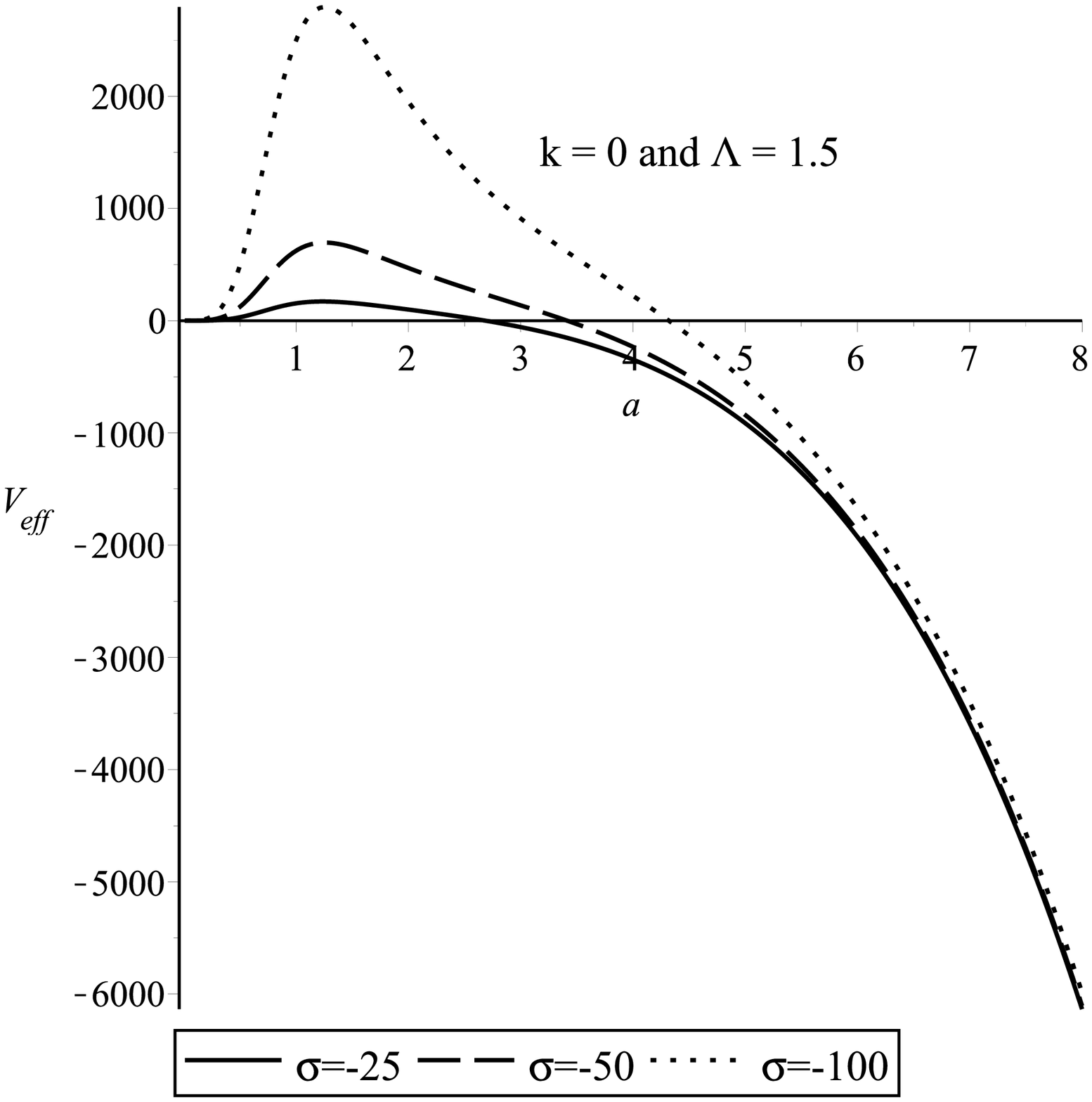}
\caption{$V_{eff}(a)$ for $k=0$ with $\Lambda = 1.5$ and different values of $\sigma$.}
\label{Figura 4} 
\end{minipage} \hfill
\end{figure}

From the total hamiltonian Eq. (\ref{Hamiltoniana}) it is possible to identify an effective potential ($V_{eff}(a)$) that comprises the terms related to the curvature of the spatial sections, cosmological constant and ad hoc potential. With the aid of Eq. (\ref{ad hoc}), $V_{eff}(a)$
is given by,
\begin{equation} 
\label{P. efetivo}
    V_{eff}(a) = 3 k a^{2} - \Lambda a^{4} + \frac{\sigma^{2} a^{4}}{(a^{3} + 1)^{2}}.
\end{equation}
Observing $V_{eff}(a)$ Eq. (\ref{P. efetivo}), it is possible to see that for all values of the parameters $\Lambda$, $\sigma$ and $k=-1$ or $k = 0$, that potential is well defined at $a=0$. In fact, it goes to zero when $a \to 0$. It is, also, possible to see that when 
$a \to \infty$ the potential $V_{eff} \to -\infty$. Another important property of $V_{eff}(a)$ Eq. (\ref{P. efetivo}), is that for all values of the parameters $\Lambda$, $\sigma$ and $k=-1$ or $k = 0$, it has only one barrier. That situation is different from the case where the curvature of the spatial section is positive ($k=1$), which was studied in Ref. \cite{germano2}. There, depending on the values of $\Lambda$ and $\sigma$, $V_{eff}(a)$ could have one or two barriers. Examples of all those properties can be seen in Figures (\ref{Figura 1}-\ref{Figura 4}).


Now, we can study the classical dynamical behavior of the model with the aid of the hamilton's equation. We may compute them from the total
hamiltonian Eq. (\ref{Hamiltoniana}), to obtain,
\begin{equation} 
\label{eq. Hamilton}
\left \{ 
\begin{array}{llllll}
      \dot{a} = & \frac{\partial N \mathcal{H}}{\partial p_{a}} = -\frac{1}{6}p_{a}, \\
			& \\
      \dot{p}_{a} = & -\frac{\partial N \mathcal{H}}{\partial a} = \frac{\partial  V_{eff}}{\partial a}, \\
			& \\
      \dot{T} = & \frac{\partial N \mathcal{H}}{\partial p_{T}} = 1, \\
			& \\
       \dot{p}_{T} = & -\frac{\partial N \mathcal{H}}{\partial T} = 0, \\
\end{array} 
\right \}
\end{equation}
where the dot means derivative with respect to the conformal time $\eta$.

One may have the general idea on how the scale factor behaves by studying the phase portraits of the models in the plane $(a,p_{a})$. Due to the fact that, as mentioned above, $V_{eff}(a)$ Eq. (\ref{P. efetivo}) for the present models have only one barrier, the phase portraits are simpler than the ones for the models with $k=+1$ \cite{germano2}. 

\begin{figure}[!htb]
\begin{minipage}[t]{0.45\textwidth}
\centering
\includegraphics[width=\linewidth]{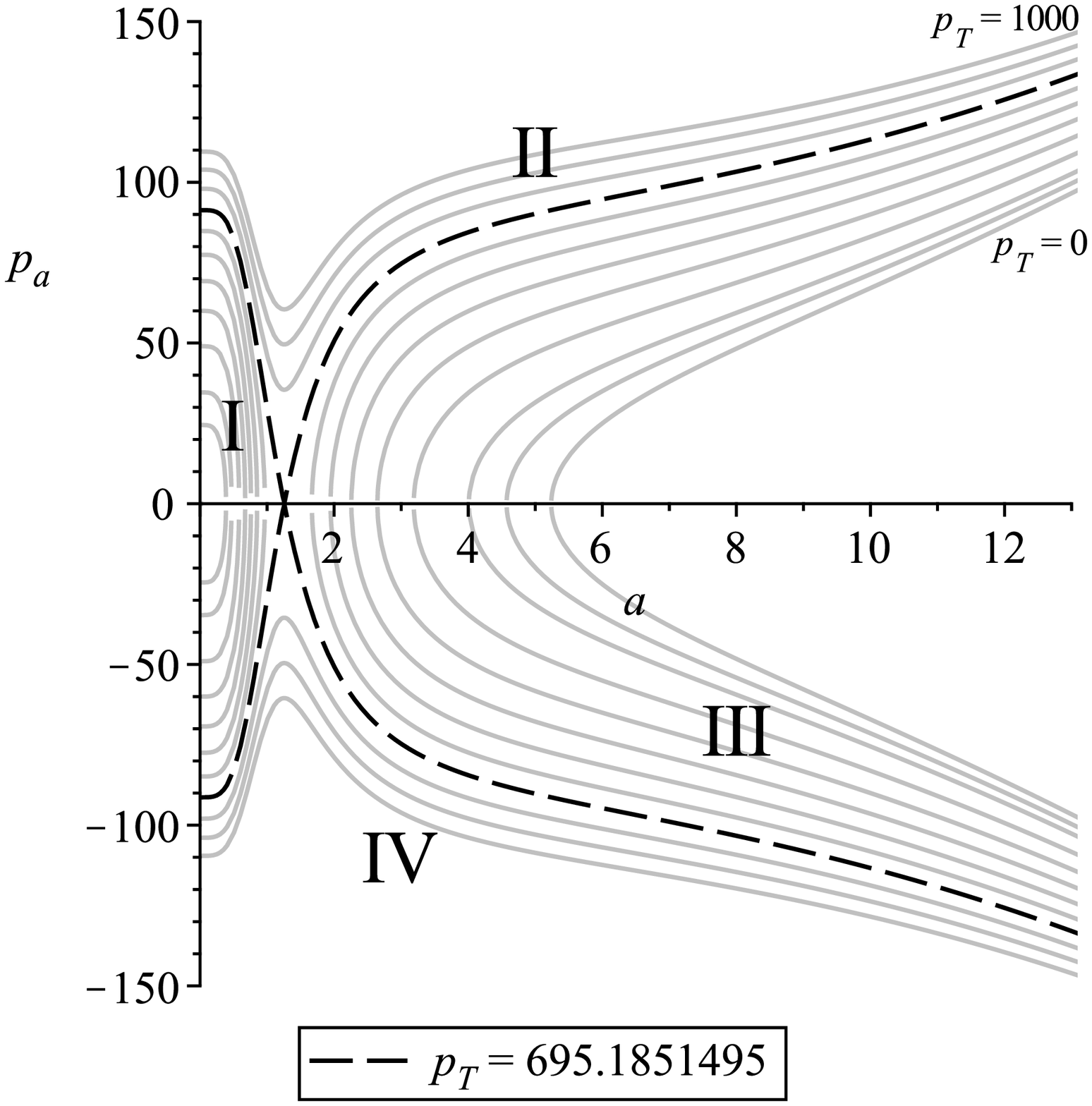}
\caption{Phase portraits in the plane $(a,p_{a})$ for the model with $k=-1$, $\Lambda = 0.01$, $\sigma = -50$ and different values of $p_T$.}
\label{Figura 5}
\end{minipage} \hfill
\begin{minipage}[t]{0.45\textwidth}
\centering
\includegraphics[width=\linewidth]{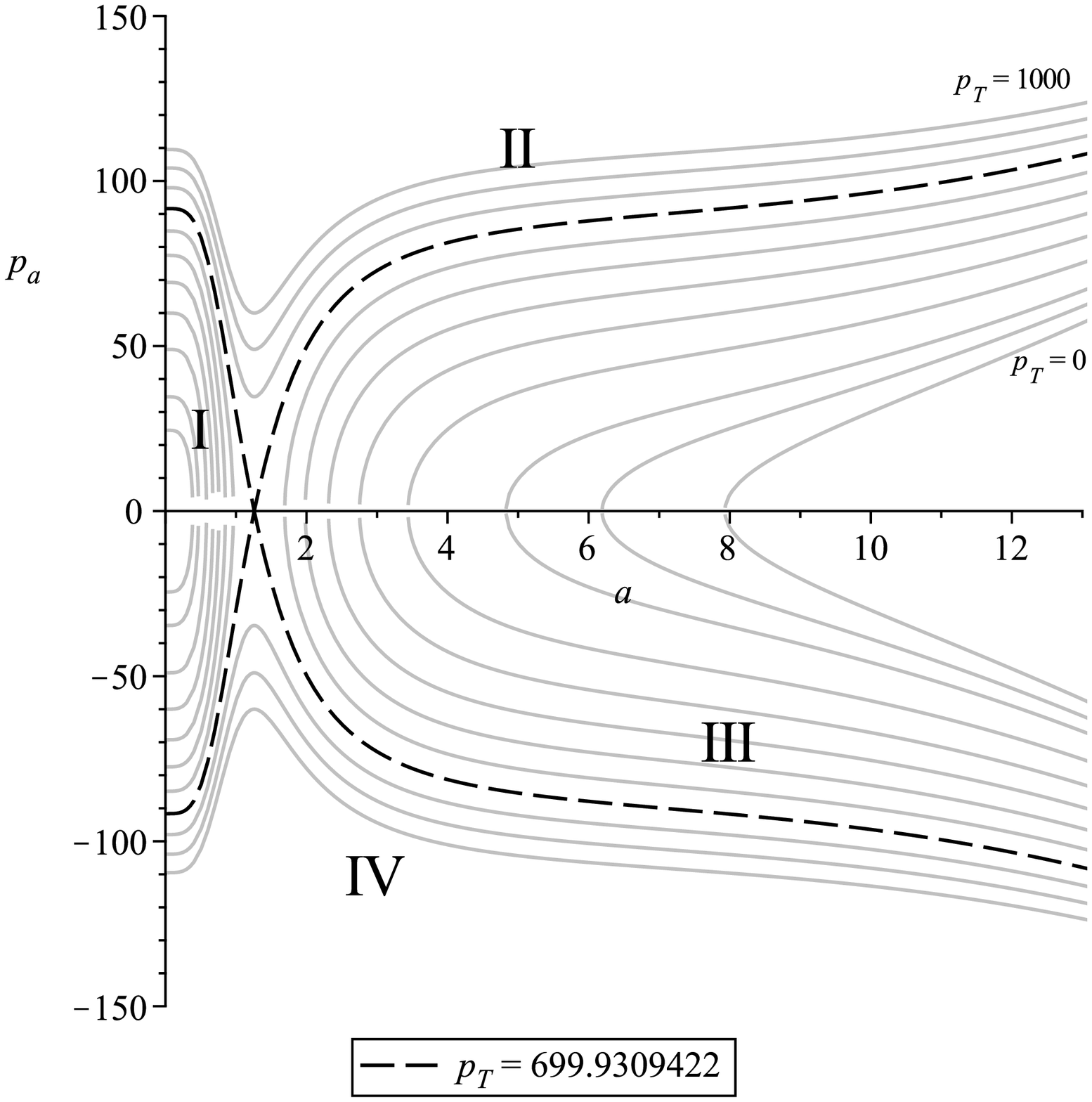}
\caption{Phase portraits in the plane $(a,p_{a})$ for the model with $k=0$, $\Lambda = 0.01$, $\sigma = -50$ and different values of $p_T$.}
\label{Figura 7}
\end{minipage} \hfill
\end{figure}
\newpage




The dashed curves in Figures \ref{Figura 5} and \ref{Figura 7} are called separatrixes. They separate different classes of solutions for a given energy $p_{T}$. Those phase portraits Figures \ref{Figura 5} and \ref{Figura 7} have, also, two fixed points, which represent stationary solutions of the model. Let us call those points $A1$ e $A2$. In particular, $A2$ is called Einstein's Universe, there the gravitational attraction and the cosmological expansion balance each other. $A1$ is located, on the plane $(a, p_a)$, by $(a = 0, p_{a} = 0)$ and energy
$p_{T} = 0$. It is the same point for Figures \ref{Figura 5} and \ref{Figura 7}. For $A2$, the points on the plane $(a, p_a)$ and the values of $p_T$ are given in Table \ref{Tabela1}.

\begin{table}[!htb]
\centering
\caption{\footnotesize Location of $A2$ for Figures \ref{Figura 5} and \ref{Figura 7}}
\label{Tabela1}
\begin{tabular}{|c | c | c|} \hline 

\footnotesize \ & \footnotesize $A2 $  & \footnotesize $p_{T} $ \\ \hline

\footnotesize Figure 5 & \footnotesize ($a$ = $1.255633946$, $p_{a} = 0) $ & \footnotesize 695.1851495\\


\footnotesize Figure 6 & \footnotesize ($a$ = $1.259875701$, $p_{a} = 0)  $ & \footnotesize 699.9309422\\

\hline
\end{tabular}
\end{table}


Observing Figures \ref{Figura 5} and \ref{Figura 7}, we may identify a first class of solutions present in the model. For $a$ and 
$p_T$ smaller than the ones for the fixed point $A_2$ and for $p_a$ greater than the ones for the fixed point $A_2$, we have
a class of solutions where the universe starts expanding from an initial Big Bang singularity, reaches a maximum size and then contracts to a final Big Crunch singularity. These solutions are located in Region I of Figures \ref{Figura 5} and \ref{Figura 7}.

Now, for $p_{a}$ and $p_T$ greater than the ones for the fixed point $A_2$, we have a second class of solutions where the universe starts expanding from an initial Big Bang singularity ($a=0$) and continues expanding to infinity values of $a$. It tends asymptotically to a De Sitter type solution. These solutions are located in Region II of Figures \ref{Figura 5} and \ref{Figura 7}.

Now, for $p_{a} < 0$ (initially), $p_T$ smaller than the ones for the fixed point $A_2$ and $a$ greater than the ones for the fixed point $A_2$, we have a third class of solutions where the universe starts contracting from an initial scale factor value, reaches a minimum size for $p_{a} = 0$ and then expands to infinity values of $a$, for $p_{a} > 0$. It tends asymptotically to a De Sitter type solution. These are the bouncing solutions for the present models. These solutions are located in Region III of Figures \ref{Figura 5} and \ref{Figura 7}.

Finally, a fourth class of solutions appears if we choose $p_{a} < 0$ and $p_T$ greater than the ones for the fixed point $A_2$. In that class of solutions the universe starts contracting from a large finite value of $a$ and continues contracting until it reaches a final Big Crunch singularity. These solutions are located in Region IV of Figures \ref{Figura 5} and \ref{Figura 7}.

The classical scale factor behavior may be computed by solving a system of ordinary differential equations. The first equation is obtained
by imposing the hamiltonian constraint $\mathcal{H}=0$ Eq. (\ref{Hamiltoniana}) and substituting, in the resulting equation, the value of $p_a$ in terms of $\dot{a}$, with the aid of Eqs. (\ref{eq. Hamilton}). That equation is the Friedmann equation for the present model and is given by,
\begin{equation} 
\label{eq. diferecial ordem 1}
    \dot{a}(0) = \pm \frac{1}{6} \sqrt{12(p_{T} - V_{eff}(a_{0}))},
\end{equation}
where $a_{0} = a(\eta=0)$ is the scale factor initial condition.
The second equation is obtained by combining the hamilton's equations $(\ref{eq. Hamilton})$, resulting in the following second order, ordinary, differential equation for $a(\eta)$,

\begin{equation} 
\label{eq. diferecial ordem 2}
    \frac{\partial^{2} a(\eta)}{\partial \eta^{2}} + k a(\eta) - \frac{2\Lambda}{3} a(\eta)^{3} + \frac{2 \sigma^{2}}{3} \frac{a(\eta)^{3}}{(a(\eta)^{3} + 1)^{2}} - \frac{\sigma^{2}a(\eta)^{6}}{(a(\eta)^{3}+1)^{3}} = 0.
\end{equation}
We solve that system of equations (\ref{eq. diferecial ordem 1}), (\ref{eq. diferecial ordem 2}), in the following way. Initially, we choose
a value for $a_0$ and substitute it in the Friedmann equation (\ref{eq. diferecial ordem 1}), in order to find the initial value for $\dot{a}$
($\dot{a}_0$). Then, we use these initial conditions in order to solve equation (\ref{eq. diferecial ordem 2}). Due to the complexity of both
equations (\ref{eq. diferecial ordem 1}), (\ref{eq. diferecial ordem 2}), we solve the system numerically. As we mentioned above, the $V_{eff}(a)$ (\ref{P. efetivo}) for both values of $k$ (-1 or 0) have just one barrier. Therefore, the results for $a(\eta)$ for both cases are very similar. Next, we solve the system of equations (\ref{eq. diferecial ordem 1}), (\ref{eq. diferecial ordem 2}), for $\Lambda=0.01$, 
$\sigma = -50$ and $k=0$ or $k=-1$, which correspond to the phase portraits shown in Figures \ref{Figura 5} and \ref{Figura 7}. For those models we find the four classes of solutions described, qualitatively, above. 

\begin{figure}[!htb]
\begin{minipage}[t]{0.45\textwidth}
\centering
\includegraphics[width=\linewidth]{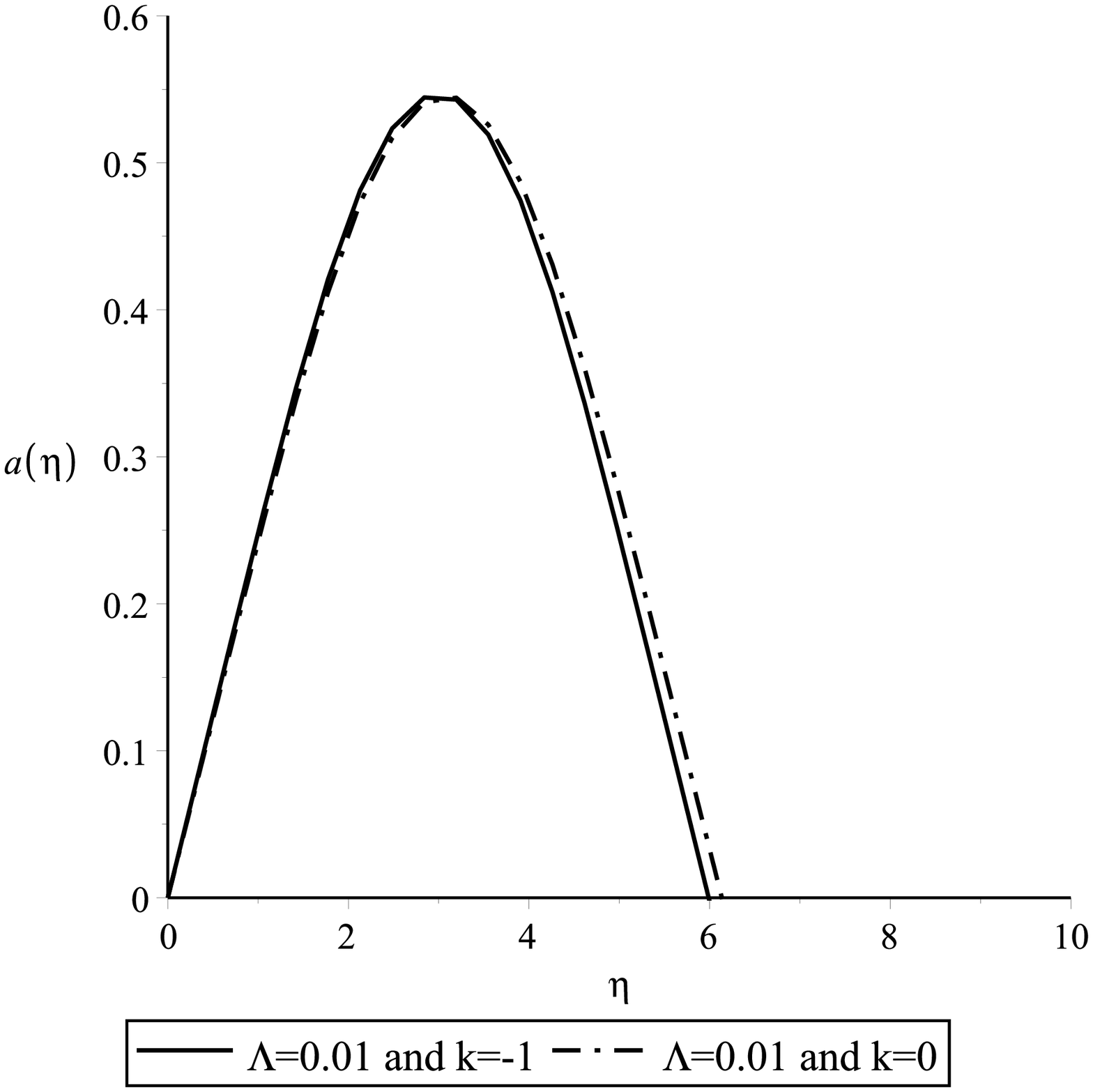}
\caption{Classical scale factor behavior for universes with $k = -1$ and $k = 0$, $\Lambda = 0.01$ and $\sigma = -50$.}
\label{Figura 9}
\end{minipage} \hfill
\begin{minipage}[t]{0.45\textwidth}
\centering
\includegraphics[width=\linewidth]{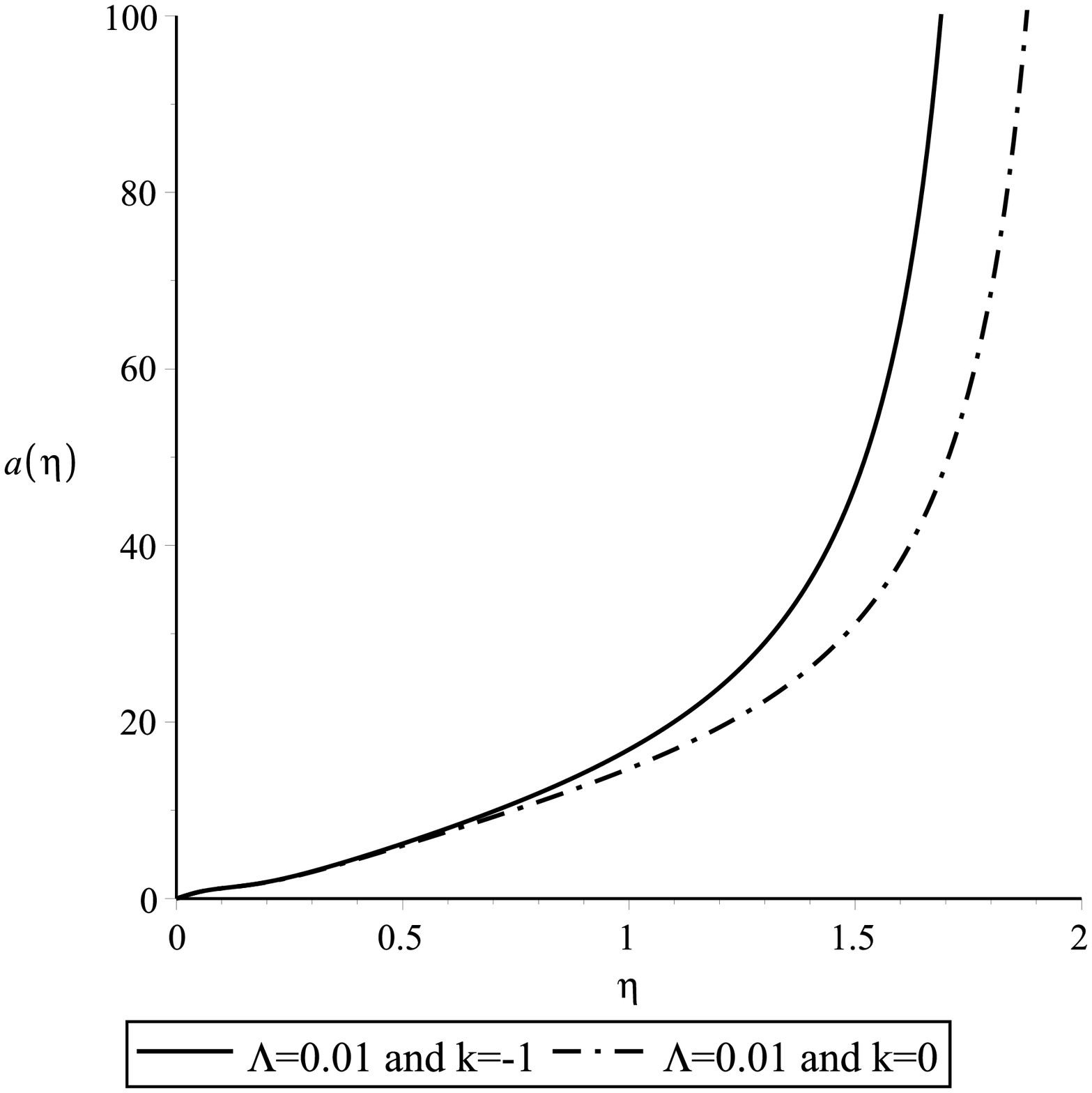}
\caption{Classical scale factor behavior for universes with $k = -1$ and $k = 0$, $\Lambda = 0.01$ and $\sigma = -50$.}
\label{Figura 10}
\end{minipage} \hfill
\begin{minipage}[t]{0.45\textwidth}
\centering
\includegraphics[width=\linewidth]{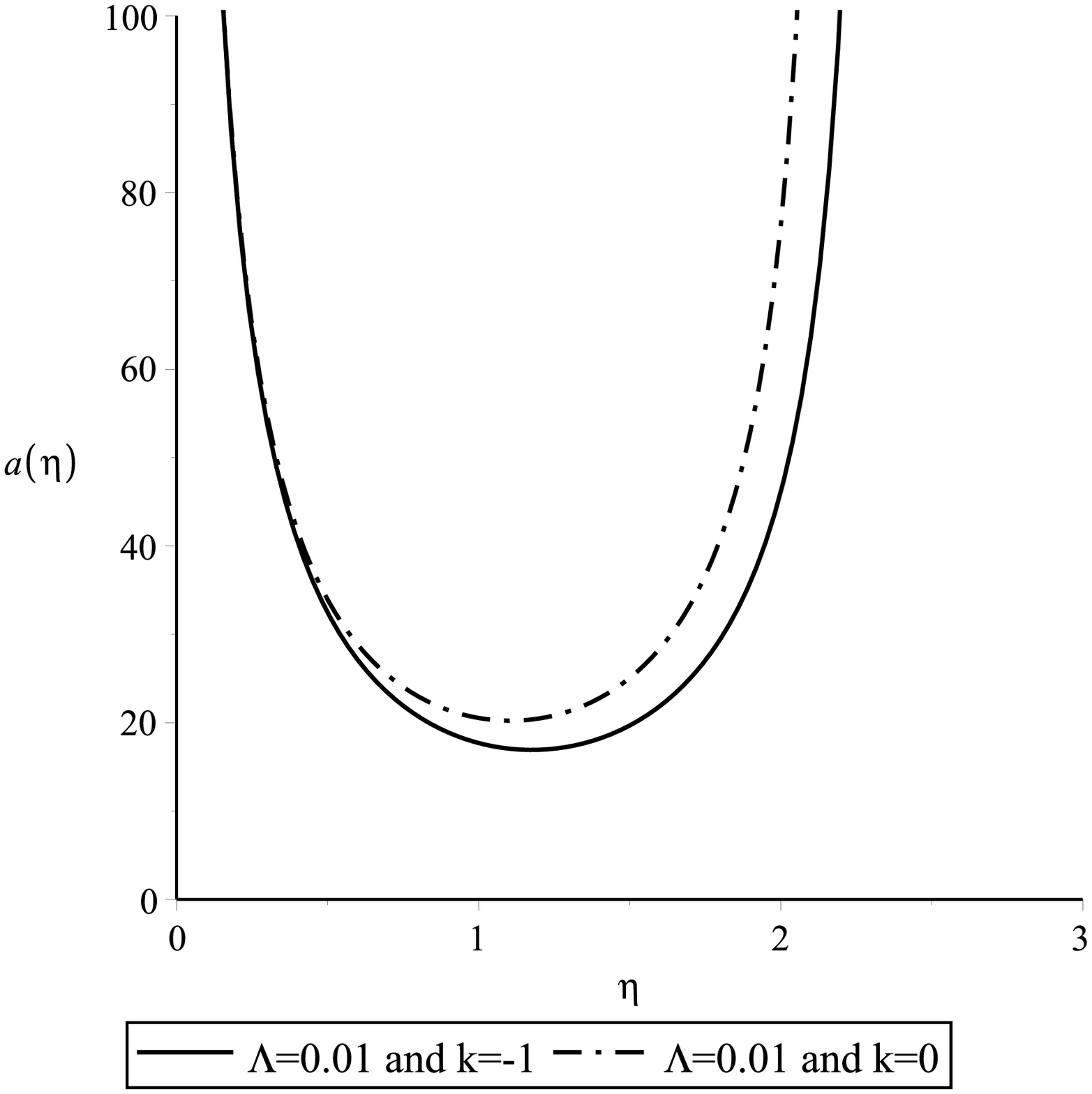}
\caption{Classical scale factor behavior for universes with $k = -1$ and $k = 0$, $\Lambda = 0.01$ and $\sigma = -50$.}
\label{Figura 11}
\end{minipage} \hfill
\begin{minipage}[t]{0.45\textwidth}
\centering
\includegraphics[width=\linewidth]{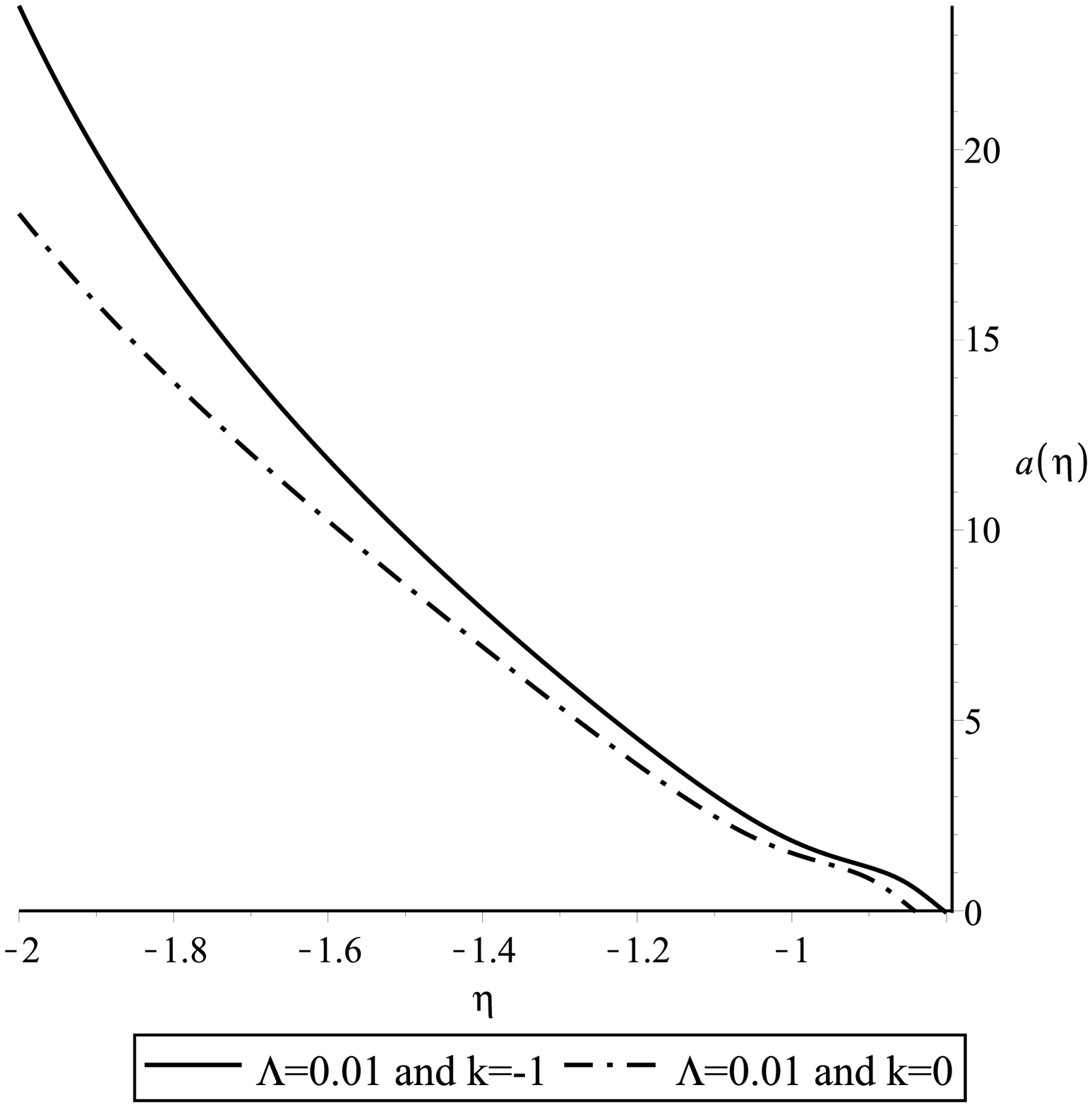}
\caption{Classical scale factor behavior for universes with $k = -1$ and $k = 0$, $\Lambda = 0.01$ and $\sigma = -50$.}
\label{Figura 12}
\end{minipage} \hfill

\end{figure}

In Figure \ref{Figura 9}, we see examples of the first class of solutions described above, for $k=-1$ and $k=0$. In order to obtain them, we
set $a(0) = 0$, $p_{T} = 164$, $\dot{a}_0 = -7.393691003$, which gives $p_{a} > 0$ from Eq. $(\ref{eq. Hamilton})$.

In Figure \ref{Figura 10}, we see examples of the second class of solutions described above, for $k=-1$ and $k=0$. In order to obtain them, we
set $a(0) = 0$, $p_{T} = 800$, $\dot{a}(0) = -16.32993162$, which gives $p_{a} > 0$ from Eq. $(\ref{eq. Hamilton})$.

In Figure \ref{Figura 11}, we see examples of the third class of solutions described above, for $k=-1$ and $k=0$. In order to obtain the solution for $k=-1$, we set $a(0) = 1000$, $p_{T} = 500$ and $\dot{a}(0) = -57743.68797$. In order to obtain the solution for $k=0$, we set 
$a(0) = 1000$, $p_{T} = 500$ and $\dot{a}(0) = -57735.02837$. We can, clearly, see from Figure \ref{Figura 11} two examples of bouncing solutions for the present models.

Finally, in Figure \ref{Figura 12}, we see examples of the fourth class of solutions described above, for $k=-1$ and $k=0$. In order to obtain the solution for $k=-1$, we set $a(0) = 10$, $p_{T} = 800$, $\dot{a}(0) = 19.79099058$, which gives $p_{a} < 0$ from Eq. $(\ref{eq. Hamilton})$.
In order to obtain the solution for $k=0$, we set $a(0) = 10$, $p_{T} = 800$, $\dot{a}(0) = 17.07873849$, which gives $p_{a} < 0$ from Eq. $(\ref{eq. Hamilton})$.

\section{Canonical Quantization, WKB Solution and WKB Tunneling Probability}
\label{CQ}

\subsection{Canonical Quantization} 
\label{cq}

In order to study the birth of the universes described by the cosmological 
models introduced in the present paper, we must quantize these models. We do that 
by using the Dirac's formalism for quantization of constrained 
systems \cite{dirac,dirac1,dirac2,dirac3}. The first step consists in introducing 
a wave-function ($\Psi$) which is a function of the canonical variables. In the 
present model these variables are $\hat{a}$ and $\hat{T}$, then,
\begin{equation}  
\label{psi}
\Psi\, =\, \Psi(\hat{a} ,\hat{T} )\, .
\end{equation}
In the second step, we demand that the operators $\hat{a}$
and $\hat{T}$ and their conjugated momenta $\hat{P}_a$ and $\hat{P}_T$,
satisfy suitable commutation relations. In the Schr\"{o}dinger picture 
$\hat{a}$ and $\hat{T}$ become multiplication operators, while their 
conjugated momenta become the following differential operators,
\begin{equation}
p_{a}\rightarrow -i\frac{\partial}{\partial a}\hspace{0.2cm},\hspace{0.2cm} 
\hspace{0.2cm}p_{T}\rightarrow -i\frac{\partial}{\partial T}\hspace{0.2cm}.
\label{momentos}
\end{equation}
In the third and final step, we impose that the operator associated to
$N \mathcal{H}$ (\ref{Hamiltoniana}) annihilates the wave-function $\Psi$ (\ref{psi}).
The resulting equation is the Wheeler-DeWitt equation for the present models.  
It resembles a time dependent, one-dimensional, Schr\"{o}dinger equation,
\begin{equation} 
\label{Wheeller-DeWitt 2}
 \Bigg{(}\frac{1}{12} \frac{\partial^{2}}{\partial a^{2}}  - 3ka^{2} + \Lambda a^{4} - 
\frac{\sigma^{2} a^{4}}{(a^{3} + 1)^{2}} \Bigg{)} \Psi(a,\tau) = -i\frac{\partial}{\partial \tau}\Psi(a,\tau)
\end{equation}
where the new variable $\tau= -T$ has been introduced.

\subsection{WKB Solution} 
\label{wkb}

Now, we want to determine the WKB approximated solution to the Wheeler-DeWitt equation (\ref{Wheeller-DeWitt 2}). We start imposing that
the solution to equation (\ref{Wheeller-DeWitt 2}) may be written as \cite{merzbacher,griffiths},
\begin{equation} 
\label{funcao de onda 1}
    \Psi(a,\tau) = \psi(a)e^{-E\tau}
\end{equation}
where $E$ is the energy associated to the radiation fluid. Introducing $\Psi(a,\tau)$
Eq. (\ref{funcao de onda 1}) in the Wheeler-DeWitt equation (\ref{Wheeller-DeWitt 2}), we obtain,
 \begin{equation} 
\label{potencial efetivo 2}
 \frac{\partial^{2}\psi(a)}{\partial a^{2}} + 12 (E - V_{eff}(a)) \psi(a) = 0,
 \end{equation}
where $V_{eff}(a)$ is given in Eq. (\ref{P. efetivo}).
Next, in Eq. (\ref{funcao de onda 1}), we consider that $\psi(a)$ is given by,
\begin{equation}
\label{wkbphase}
\psi(a) = A(a) e^{i\phi(a)}, 
\end{equation}
where $A(a)$ is the amplitude and $\phi(a)$ is the phase. Introducing $\psi(a)$ Eq. (\ref{wkbphase}) in Eq. (\ref{potencial efetivo 2}) and supposing that the amplitude $A(a)$ varies slowly as a function of $a$, we find the following general solutions for Eq. (\ref{potencial efetivo 2}):

$(i)$ For regions where $E > V_{eff}(a)$,
\begin{equation}
\label{generalwkb}
\psi(a) = \frac{C}{\sqrt{K(a)}} e^{\pm\frac{i}{\hbar}\int K(a)da},
\end{equation}
where $C$ is a constant and
\begin{equation}
\label{wkbK} 
K(a) = \sqrt{12(E - V_{eff}(a))}.
\end{equation}

$(ii)$ For regions where $E < V_{eff}(a)$,
\begin{equation}
\label{generalwkb1}
\psi(a) = \frac{C1}{\sqrt{k(a)}} e^{\pm\frac{1}{\hbar}\int k(a)da},
\end{equation}
where $C1$ is a constant and
\begin{equation}
\label{wkbk}
k(a) = \sqrt{12(V_{eff}(a) - E)}.
\end{equation}

\subsection{WKB Tunneling Probability} 
\label{tpwkb}

Finally, using those WKB solutions, we want to determine the quantum mechanical tunneling probabilities for the birth of the present universes. More precisely, the probabilities that the present universes will tunnel through $V_{eff}$. An important condition for the
tunneling process is that the energy $E$, of the wavefunction, be smaller than the maximum value of $V_{eff}(a)$. If we impose that condition,
we may divide the $a$ axis in three distinct regions with respect to the points where $E$ intercepts $V_{eff}(a)$ (\ref{P. efetivo}), which are: (1) Region I - It extends from the origin until the point where $E$ intercepts $V_{eff}(a)$ at the left ($a_{l}$), $0 < a < a_{l}$; (2) Region II - It extends from the point where $E$ intercepts $V_{eff}(a)$ at the left until the point where $E$ intercepts $V_{eff}(a)$ at the right ($a_{r}$), $a_{l} < a < a_{r}$. That region is entirely inside $V_{eff}(a)$; (3) Region III - It extends from the point where $E$ intercepts $V_{eff}(a)$ at the right until the infinity, $a_{r} < a < \infty$. Now, we may write the WKB solutions Eqs.
(\ref{generalwkb}) and (\ref{generalwkb1}) for each one of these three regions,
\begin{eqnarray} 
\label{wkbregions}
\psi(a) & = & \frac{A}{\sqrt{K(a)}} e^{i\int_{a}^{a_l}K(a)da} + \frac{B}{\sqrt{K(a)}} e^{-i\int_{a}^{a_l}K(a)da} \quad \mathrm{I} \quad (0 < a < a_{l})\nonumber\\
\psi(a) & = & \frac{C}{\sqrt{k(a)}} e^{-\int_{a_{l}}^{a_r}k(a)da} + \frac{D}{\sqrt{k(a)}} e^{\int_{a_{l}}^{a_r}k(a)da} \quad  \mathrm{II} \quad (a_{l} < a < a_{r})\nonumber\\
\psi(a) & = & \frac{F}{\sqrt{K(a)}} e^{i\int_{a_{r}}^{a}K(a)da} + \frac{G}{\sqrt{K(a)}} e^{-i\int_{a_{r}}^{a}K(a)da} \quad \mathrm{III} \quad (a_{r} < a < \infty)\nonumber\\
\end{eqnarray}
where $A,B,C,D,E,F,G$ are constant coefficients to be determined. One may establish a relationship between all these coefficients 
$A,B,C,D,E,F,G$ with the aid of the connections formulas, which are important formulas of the WKB approximation \cite{merzbacher,griffiths}. The relationship is given by the following equation,
\begin{equation}
\label{matrix}
\left(\begin{array}{c}
A \\ B
\end{array} \right)
= \frac{1}{2}   
\left(\begin{array}{cc} 
2\theta + \frac{1}{2\theta} & i(2\theta - \frac{1}{2\theta}) \\ 
-i(2\theta - \frac{1}{2\theta}) &  2\theta + \frac{1}{2\theta}
\end{array} \right)
\left(\begin{array}{c}
F \\ G
\end{array} \right),
\end{equation}
where $\theta$ is given by,
\begin{equation} 
\label{theta}
    \theta = e^{\int_{a_{l}}^{a_{r}} k(a) da} = e^{\int_{a_{e}} ^{a_{d}} da \sqrt{12 (3ka^{2} - \Lambda a^{4} + \frac{\sigma^{2} a^{4}}{(a^{3} + 1)^{2}} - E)}}.
\end{equation}

Let us consider, now, that the incident wavefunction ($\psi_{inc}$) with energy $E$ propagates from the origin to the left of $V_{eff}(a)$ in Region I. When the wavefunction reaches $V_{eff}(a)$ at $a_l$, part of the incident wavefunction is reflected back to Region I and part tunnels through $V_{eff}(a)$ in Region II. When the wavefunction emerges from $V_{eff}(a)$ at $a_r$, it produces a transmitted component 
($\psi_{trans}$) which propagates to infinity in Region III. By definition the tunneling probability ($TP_{WKB}$) is given by,
\begin{equation} 
\label{PTwkb}
    TP_{WKB} = \frac{ |\psi_{trans} \sqrt{k_{trans}}|^{2}}{ |\psi_{inc} \sqrt{k_{inc}}|^{2}} = \frac{  | F  |^{2}}{ | A |^{2}},
\end{equation}
where we are assuming that there is no incident wavefunction from the right, it means that $G = 0$ in Eq. (\ref{matrix}). With the aid
of Eq. (\ref{matrix}), $TP_{WKB}$ becomes,
\begin{equation} 
\label{PTwkb2}
     TP_{WKB} = \frac{4}{(2\theta + \frac{1}{2\theta})^{2}}.
\end{equation}

\section{Results}
\label{results}

Now, we want to quantitatively compute the tunneling probabilities for the birth of the universes described by the present models. 
These tunneling probabilities are measured by $TP_{WKB}$ (\ref{PTwkb2}). They depend on: (i) the radiation energy $E$, (ii) the cosmological constant $\Lambda$ and (iii) the ad hoc potential parameter $\sigma$.

\subsubsection{$TP_{WKB}$ as a function of $E$}

\begin{figure}[!htb]
\begin{center}
    \includegraphics[width=0.6\linewidth]{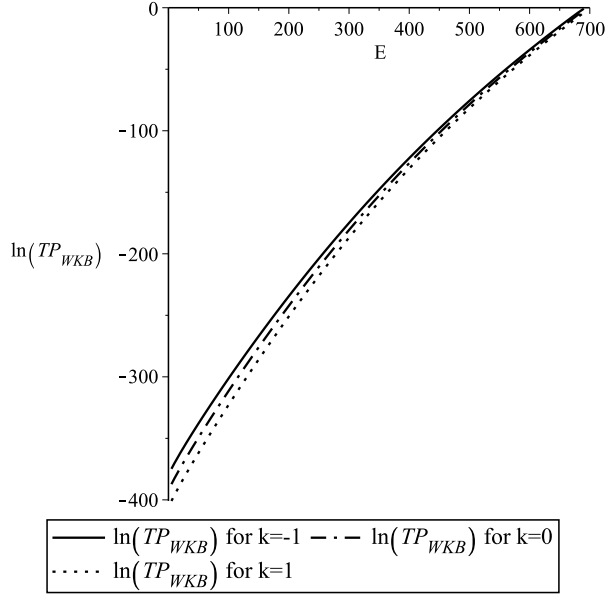}
\end{center}
\caption{WKB Tunneling Probabilities as functions of the energy $E$, for $\sigma = -50$ and $\Lambda = 1.5$. Each curve corresponds to a different value of the spatial curvature $k$.}
\label{Figura 13}
\end{figure}

If we fix the values of $\Lambda$, $\sigma$ and $k$, $TP_{WKB}$ Eq. (\ref{PTwkb2}) becomes a function of the energy $E$. In order to determine how that tunneling probability depends on $E$, we compute $TP_{WKB}$ Eq. (\ref{PTwkb2}) for 70 different values of $E$ with $\sigma = -50$ and $\Lambda = 1.5$. As a matter of completeness and in order to facilitate the comparison, we shall compute the values for the models with $k=1$ besides the ones for models with $k=-1,0$. Therefore, we repeat those calculations three times, one for each value of $k$. We choose values of $E$, such that, they are smaller than the maximum barrier value ($V_{effmax}$). For $k=-1$ $V_{effmax}=691.5188154$, for 
$k=0$ $V_{effmax}=696.2063154$ and for $k=1$ $V_{effmax}=700.8938154$. The energies are given by: $E = \{E_{1}=5,E_{2}=10,E_{3}=20, ...,E_{68}= 670, E_{69}= 680,  E_{70}=690\}$. 
The curves $\ln(TP_{WKB})$ versus $E$, for each $k$, are given in Figure \ref{Figura 13}. We use the natural logarithm of $TP_{WKB}$ because some values of that tunneling probability are very small. Observing Figure \ref{Figura 13}, we notice that $TP_{WKB}$ increases for greater values of $E$. Thus, it is more likely that the universe is born with the greatest value of the radiation energy $E$. From Figure \ref{Figura 13}, we also notice that $TP_{WKB}$ is greatest for $k=-1$, decreases for $k=0$ and decreases even further for $k=1$. So, it is more likely that the universe is born with negatively curved spatial sections. 

\subsubsection{$TP_{WKB}$ as a function of $\Lambda$}

\begin{figure}[!htb]
\begin{center}
    \includegraphics[width=0.6\linewidth]{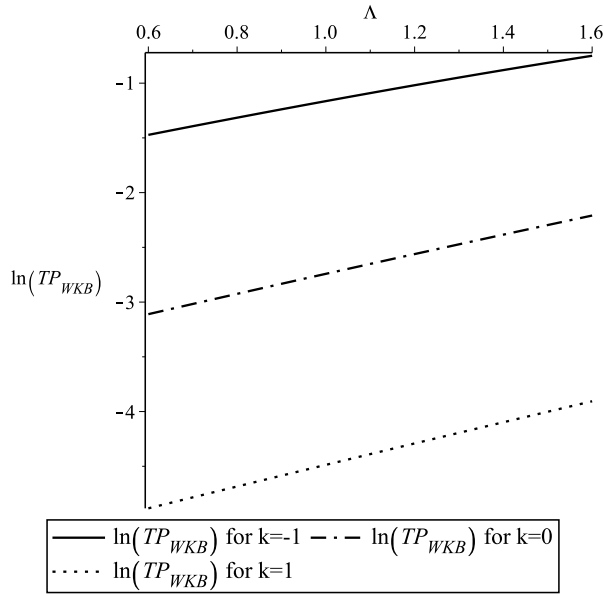}
\end{center}
\caption{WKB Tunneling Probabilities as functions of the cosmological constant $\Lambda$, for $\sigma = -50$ and $E = 690$. Each curve corresponds to a different value of the spatial curvature $k$.}
\label{Figura 14}
\end{figure}

If we fix the values of $E$, $\sigma$ and $k$, $TP_{WKB}$ Eq. (\ref{PTwkb2}) becomes a function of the cosmological constant $\Lambda$. In order to determine how that tunneling probability depends on $\Lambda$, we compute $TP_{WKB}$ Eq. (\ref{PTwkb2}) for 21 different values of $\Lambda$ with $\sigma = -50$ and $E = 690$. We repeat those calculations three times, one for each value of $k$. We choose values of $\Lambda$, such that, $E = 690$ is always smaller than $V_{effmax}$. The cosmological constant values are given by: $\Lambda = \{\Lambda_{1} = 0.6, \Lambda_{2} = 0.65, \Lambda_{3} = 0.7, ... , \Lambda_{19} = 1.5, \Lambda_{20} = 1.55, \Lambda_{21} = 1.6\} $. 
The curves $\ln(TP_{WKB})$ versus $\Lambda$, for each $k$, are given in Figure \ref{Figura 14}. We use the natural logarithm of $TP_{WKB}$ because some values of that tunneling probability are very small. Observing Figure \ref{Figura 14}, we notice that $TP_{WKB}$ increases for greater values of $\Lambda$. Therefore, it is more likely that the universe is born with the greatest value of $\Lambda$. From Figure \ref{Figura 14}, we also notice that $TP_{WKB}$ is greatest for $k=-1$, decreases for $k=0$ and decreases even further for $k=1$. Thus, it is more likely that the universe is born with negatively curved spatial sections. 

\subsubsection{$TP_{WKB}$ as a function of $\sigma$}

\begin{figure}[!htb]
\begin{center}
    \includegraphics[width=0.6\linewidth]{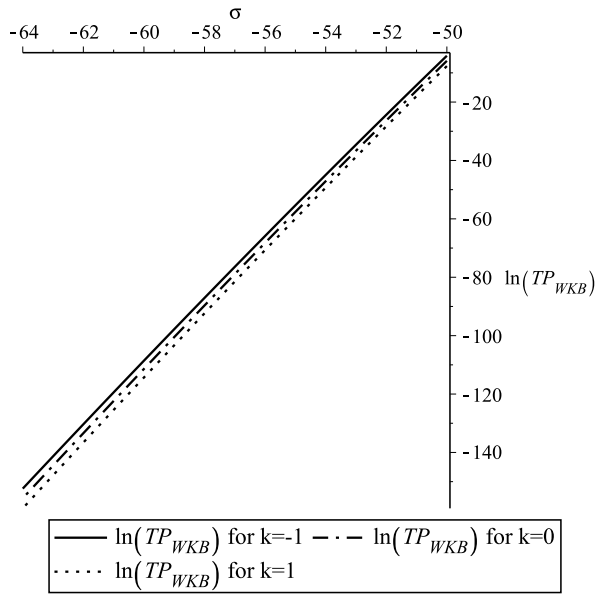}
\end{center}
\caption{WKB Tunneling Probabilities as functions of the ad hoc potential parameter $\sigma$, for $\Lambda = 1.5$ and $E = 680$. Each curve corresponds to a different value of the spatial curvature $k$.}
\label{Figura 15}
\end{figure}

If we fix the values of $E$, $\Lambda$ and $k$, $TP_{WKB}$ Eq. (\ref{PTwkb2}) becomes a function of the ad hoc potential parameter $\sigma$. In order to determine how that tunneling probability depends on $\sigma$, we compute $TP_{WKB}$ Eq. (\ref{PTwkb2}) for 29 different values of $\sigma$ with $\Lambda = 1.5$ and $E = 680$. We repeat those calculations three times, one for each value of $k$. We choose values of $\sigma$, such that, $E = 680$ is always smaller than $V_{effmax}$. The ad hoc potential parameter values are given by: $\sigma = \{ \sigma_{1}= -50, \sigma_{2}= -50.5, \sigma_{3}= -51, ..., \sigma_{27}= -63, \sigma_{28}= -63.5, \sigma_{29}= -64  \}$. 
The curves $\ln(TP_{WKB})$ versus $\sigma$, for each $k$, are given in Figure \ref{Figura 15}. We use the natural logarithm of $TP_{WKB}$ because some values of that tunneling probability are very small. Observing Figure \ref{Figura 15}, we notice that $TP_{WKB}$ decreases for greater absolute values of $\sigma$. Therefore, it is more likely that the universe is born with the smallest possible absolute value of $\sigma$. From Figure \ref{Figura 15}, we also notice that $TP_{WKB}$ is greatest for $k=-1$, decreases for $k=0$ and decreases even further for $k=1$. So, it is more likely that the universe is born with negatively curved spatial sections. 

\section{Conclusions}
\label{CC}

In this work we studied the birth of FLRW models 
with zero ($k=0$) and negative ($k=-1$) curvatures of the spatial sections.
The model with $k=1$ was studied in Ref. \cite{germano2}. Here, we considered few 
results obtained in Ref. \cite{germano2} in order to compare them with the new results obtained 
for the models with $k=0$ and $k=-1$. The material content of the models 
is composed of a radiation perfect fluid and a positive cosmological constant. The models also 
have the presence of an ad hoc potential which origin is believed to be of geometrical nature.
At the classical level, we studied the models by drawing phase portraits in the plane ($a$, $p_a$).
We identified all possible types of solutions, including some new bouncing solutions. We explicitly, 
solved the Einstein's equations and gave examples of all possible types of classical solutions.

In order to describe the birth of these universes, we quantized them using quantum cosmology. 
Initially, we obtained the Wheeler-DeWitt equations and solved them using the WKB approximation. 
We notice that the presence of $V_{ah}$ produces a barrier for any value of $k$. It means that 
we may describe the birth of the universe through a tunneling mechanism, for any curvature of the 
spatial sections, not only for the usual case $k=1$. We, explicitly, computed the tunneling 
probabilities for the birth of the different models of the universe, as functions of the radiation
energy $E$, the cosmological constant $\Lambda$ and the ad hoc potential parameter $\sigma$. We 
compared the WKB tunneling probability behavior for different values of $k$.

From our results, we noticed that $TP_{WKB}$ increases for greater values of $E$. Therefore, it is 
more likely that the universe is born with the greatest value of the radiation energy $E$.
We, also, noticed that $TP_{WKB}$ increases for greater values of $\Lambda$. Thus, it is more 
likely that the universe is born with the greatest value of $\Lambda$.
We, also, noticed that $TP_{WKB}$ decreases for greater absolute values of $\sigma$. Hence, it 
is more likely that the universe is born with the smallest possible absolute value of $\sigma$.
In all models we have studied, we noticed that $TP_{WKB}$ is greatest for $k=-1$, decreases for $k=0$ 
and decreases even further for $k=1$. So, it is more likely that the universe is born with 
negatively curved spatial sections.


{\bf Acknowledgments}. D. L. Canedo thanks Coordena\c{c}\~{a}o de\\ Aperfei\c{c}oamento de Pessoal de N\'{i}vel Superior (CAPES) and Universidade Federal de Juiz de Fora (UFJF) for his scholarships. G. A. Monerat thanks FAPERJ for financial support and Universidade do Estado do Rio de Janeiro (UERJ) for the Proci\^{e}ncia grant.

\appendix 

\section{Radiation fluid hamiltonian}
\label{SF}

In the present model, the starting point of the Schutz formalism is the description of the fluid four-velocity $U_{\nu}$ in terms of the potentials $\mu$, $\phi, \theta$ and $S$,
\begin{equation}
\label{potencial}
U_{\nu} = \frac{1}{\mu}(\phi_{,\nu} +  \theta S_{,\nu}) \, ,
\end{equation}
where $\mu$ is the specific enthalpy, $S$ is the specific entropy and the potentials $\phi$ and $\theta$ have no clear physical meaning.
The four-velocity is subjected to the normalization condition,
\begin{equation}
\label{normalization}
U^{\nu}U_{\nu} = -1
\end{equation}
In what follows, we will use the following thermodynamic equations, 
\begin{equation}  
\label{eqTermodinâmica}
   \rho = \rho_{0} (1 + \Pi),\qquad
   \mu = (1 + \Pi) + \frac{p}{\rho_{0}},\qquad
   TdS = d\Pi + pd\left(\frac{1}{\rho_{0}}\right),
\end{equation}
where $\Pi$ is the specific internal energy, $T$ is the absolute temperature and $\rho_{0}$ is the rest mass density. Combining those equations, we may write,
\begin{equation} 
\label{eqTermodinâmica2}
   T = 1 + \Pi, \qquad
   S = \ln(1 + \Pi) \frac{1}{\rho_{0}^{\frac{1}{3}}}
\end{equation}
Now, we can write the specific enthalpy $\mu$ in terms of the other thermodynamic potentials presents in Eq.(\ref{potencial}), with the aid of the normalization condition Eq.(\ref{normalization}),
\begin{equation} 
\label{entalpia}
    \mu = \frac{1}{N}(\dot{\phi} + \theta \dot{S}).
\end{equation}
If we combine the Eqs. (\ref{eqTermodinâmica}), (\ref{eqTermodinâmica2}) and (\ref{entalpia}), we may write the radiation energy density as,
\begin{equation} 
\label{rho}
    \rho = \Bigg{(}\frac{\frac{1}{N}(\dot{\phi} + \theta \dot{S})}{\frac{4}{3}}\Bigg{)}^{4} e^{-3S}
\end{equation}
Introducing the above expression of $\rho$ Eq.(\ref{rho}) in the radiation fluid action Eq.(\ref{Ação fluido}), we find with the aid of 
Eq.(\ref{eqstate}),
\begin{equation} 
\label{Ação fluido Apêndice (2)}
\int_{M}d^{4}x\sqrt{-g} \frac{1}{3} \rho_{rad} = \int_{M}d^{4}x\sqrt{-g} \frac{1}{3}\Bigg{(}\frac{\frac{1}{N}(\dot{\phi} + \theta \dot{S})}{\frac{4}{3}}\Bigg{)}^{4} e^{-3S},
\end{equation}
Next, we identify from the radiation fluid action Eq.(\ref{Ação fluido Apêndice (2)}) its lagrangian $L_{f}$,
\begin{equation} 
\label{lagrangiana fluido}
    L_{f} = \frac{27}{256} \frac{ a^{3}}{N^{3}} {(\dot{\phi} + \theta \dot{S})}^{4} e^{-3S}
\end{equation}
From that lagrangian, we compute the canonically conjugated momenta to the canonical variables $\phi$ ($p_{\phi}$) and $S$ ($p_{S}$), in the usual way,
\begin{equation}  
\label{momentos canônicos fluido}
    p_{\phi} = \frac{\partial L_{f} }{\partial \dot{\phi}} = \frac{27}{64} \frac{ a^{3}}{N^{3}} {(\dot{\phi} + \theta \dot{S})}^{3} e^{-3S},
    \qquad
     p_{S} = \frac{\partial L_{f} }{\partial \dot{S}} = \theta   p_{\phi}
\end{equation}
The general expression for the fluid total hamiltonian $N \mathcal{H}_{f}$, in the present model, is given by,
\begin{equation} 
\label{hamiltoniana fluido}
N \mathcal{H}_{f} = \dot{\phi} p_{\phi} + \dot{S} p_{S} - N L_{f},
\end{equation}
Introducing the fluid lagrangian Eq.(\ref{lagrangiana fluido}) and the canonically conjugated momenta Eq.(\ref{momentos canônicos fluido}) in the fluid total hamiltonian expression Eq.(\ref{hamiltoniana fluido}), we find,
\begin{equation} 
\label{hamiltoniana fluido_1}
   N \mathcal{H}_{f} = \frac{ {p_{\phi}}^{\frac{4}{3}}}{a} e^{S}.
\end{equation}
We may greatly simplify the fluid total hamiltonian expression Eq.(\ref{hamiltoniana fluido_1}) by performing the following canonical transformations 
\cite{rubakov},
\begin{equation} 
\label{transformação canônica}
    T = p_{s}e^{-S}{p_{\phi}}^{-\frac{4}{3}},
		\qquad
    p_{T} = {p_{\phi}}^{\frac{4}{3}}e^{S},
		\qquad
    \bar{\phi} = \phi - \frac{4}{3} \frac{p_{S}}{p_{\phi}},
		\qquad
    \bar{p}_{\phi} = p_{\phi}.    
\end{equation}
If we rewrite the fluid total hamiltonian Eq.(\ref{hamiltoniana fluido_1}) in terms of the new canonical variables and their conjugated momenta Eqs.(\ref{transformação canônica}), we obtain,
\begin{equation} 
\label{hamiltoniana fluido_2}
   N\mathcal{H}_{f} = \frac{P_{T}}{a}.
\end{equation}
Observing that last equation, we notice that the canonical variable $T$, associated to the radiation fluid, will play the role of time in the quantum version of those models.


















\end{document}